# Magnetic order of Mn-doped ZnO:
# A Monte Carlo simulation of Carriers Concentration's effect

Lalla Btissam Drissi[1, 2]

1- INANOTECH (Institut of Nonmaterials and Nanotechnology)- MASCIR
(Moroccan Foundation for Advanced Science, Innovation and Research)
Rabat, Morocco
2- International Centre for Theoretical Physics, ICTP, Trieste, Italy,

ldrissi@ictp.it, b.drissi@mascir.com

## Abstract

Recently, diluted Magnetic Semiconductors (DMS) doped with a small concentration of magnetic impurities, especially DMS doped with transition metal, inducing ferromagnetic DMSs have attracted a great interest. Using the Monte Carlo method within the Ising model, we study the magnetic properties of doped Mn ions in semi-conductor for different carrier's concentration. For the case of $Zn_{1-x}Mn_xO$, the results of our calculations show the effect of carriers in understanding the existence and the control of the magnetic order. We give also the exact values of carriers' concentration that should be adopted or avoided in order to get ferromagnetic phase with high Curie temperature.

**Keywords:** DMS, ZnO, Magnetic property, MC simulations, RKKY, Ising Model, Curie Temperature, Carrier mediated ferro

## Introduction

Recently, a great interest has been devoted to Diluted Magnetic Semiconductors (DMS) doped with a small concentration of magnetic impurities inducing ferromagnetic DMSs. In particular, DMS based on III-V and II-VI semiconductors doped with transition metal are deeply investigated by both theoretical and experimental scientists in order to use them for spintronic devices such as spin-valve transistor, spin light emitting diodes, optical isolator, non-volatile memory. Based on a previous work **[1]**, we present in this poster, using the Monte Carlo method within the Ising model, the magnetic properties of doped Mn ions in semi-conductor for different carrier's concentration. For the case of $Zn_{1-x}Mn_xO$, the results of our calculations show the effect of carriers in understanding the existence and the control of the magnetic order. We give also the exact values of carriers' concentration that should be adopted or avoided in order to get ferromagnetic phase with high Curie temperature.

# The Model

ZnO has wurtzite crystal structure (P63mc), where each atom of Zinc is surrounded by four cations of oxygen at the corners of a tetrahedron and vice versa. This tetrahedral coordination is typical of sp3 covalent bonding and some of the divalent sites Zn substituted by the magnetic ions Mn.

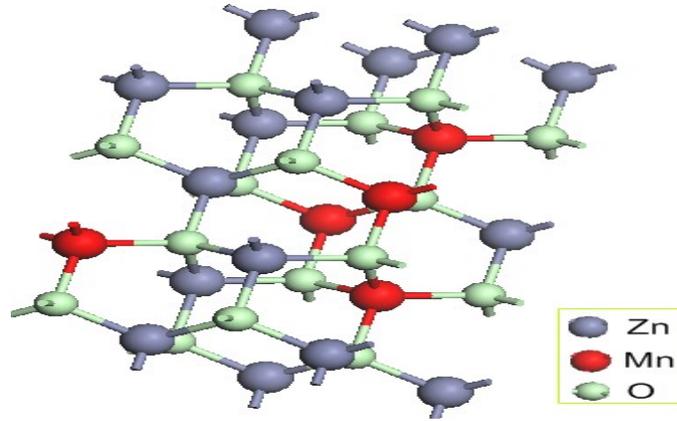

Fig1: The wurtzite structure of ZnO doped by some magnetic ions Mn

# The Hamiltonian of the Model

The system, namely ZnO doped with different concentrations of Mn, is described by the following Hamiltonian

$$H = \sum_{i,j} J(r_{ij}) \left( n_i n_j S_i^z S_j^z \right) \quad \text{for } i \neq j$$

where $r_{ij}$ is the separation between moments at the two sites $i$ and $j$ in the hexagonal structure and $n_i$ is the magnetic impurity occupation number and $J(r_{ij}) = J^{RKKY}(r)$ is the well-known RKKY range function. Notice that for the three distances D1, D2 and D3 corresponding respectively to the first, second and third nearest neighbors, the function $J(r_{ij})$ vary as follows,

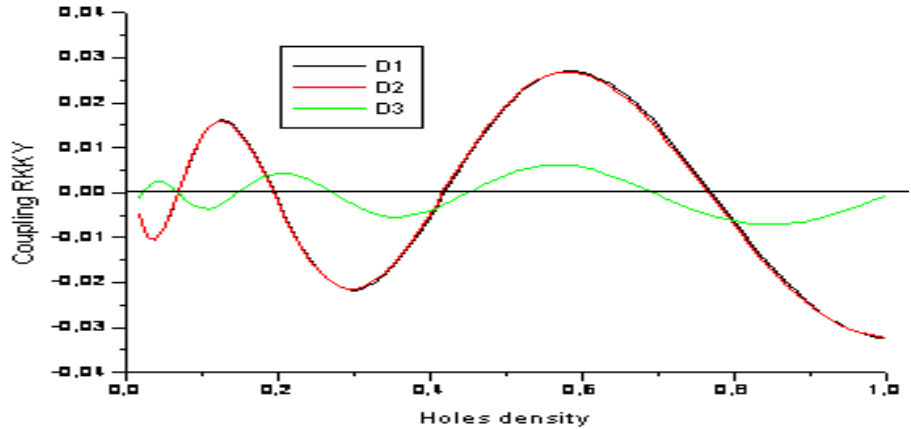

Fig 2: The variation of the RKKY function $J(r_{ij})$ in terms of hole density $n_C$ for the three distances D1, D2 and D3 of the third first nearest neighbors

# The Main Results

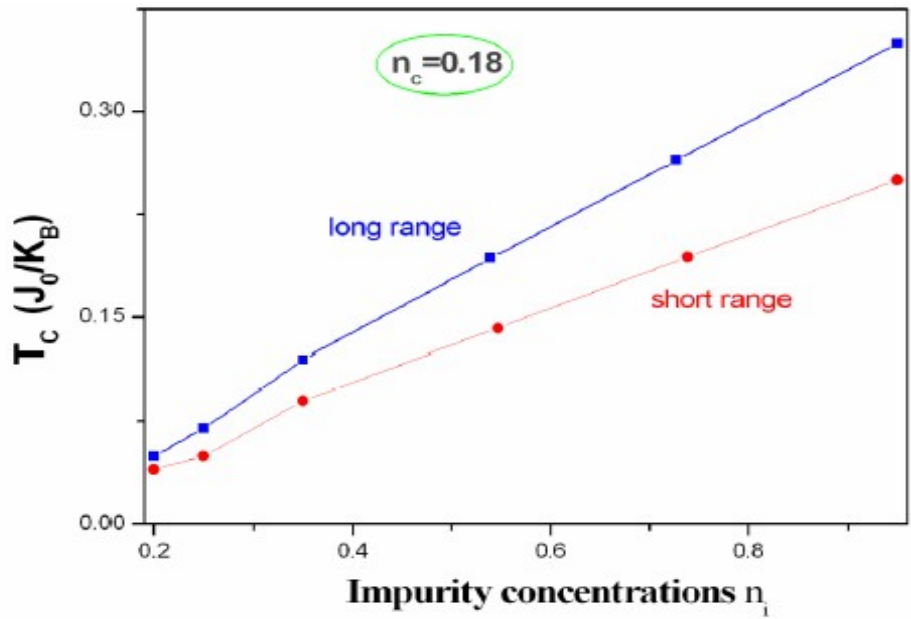

Figure 3 : At fixed $n_c = 0.18$, the Curie temperature versus the impurity concentrations $n_i$ of Mn for the short and long range case corresponding respectively to damping scale D1 and D3, shows: NO DEVIATION FROM STRICTLY LINEAR BEHAVIOR

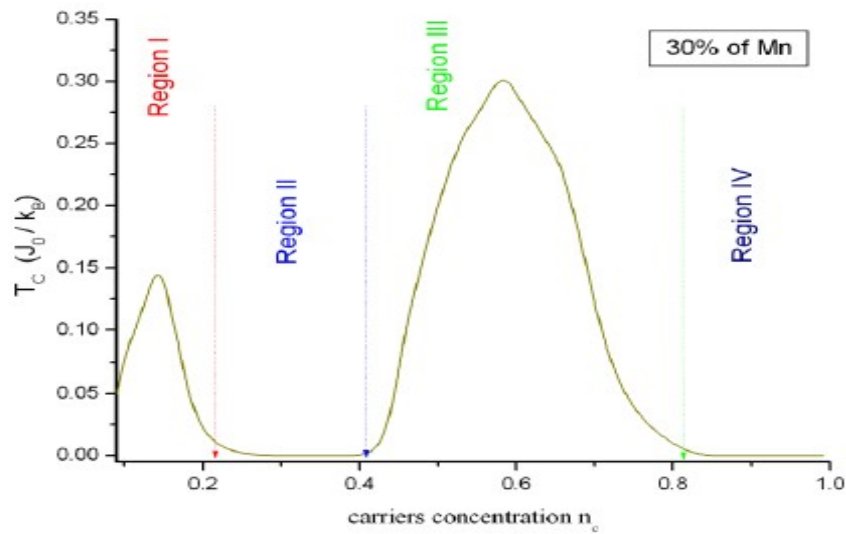

Figure 4 : For the specific Mn concentration 30%, the transition temperature, $T_C$ as function of the carriers concentration, $n_c$, reveals that we have four main different regions: REGION I & III giving FM phase with different values of $T_C$, however REGION II & IV must be avoided.

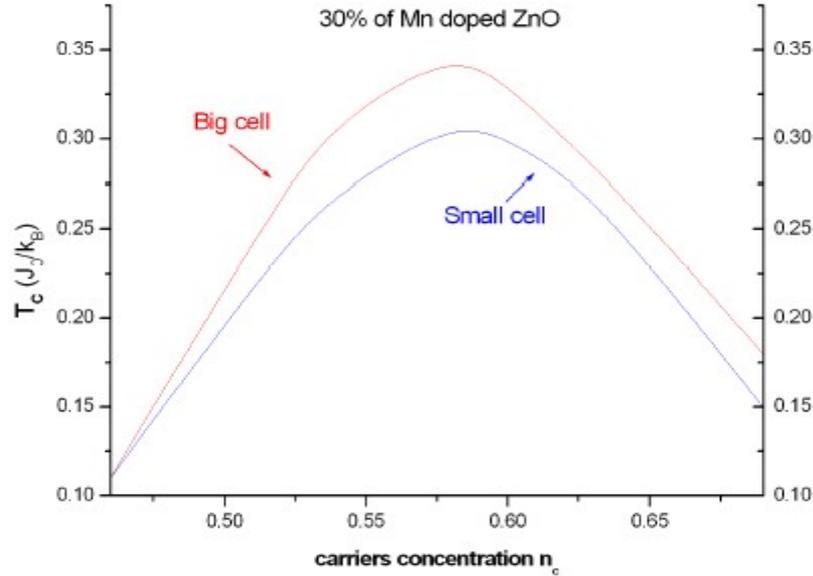

Figure 5: Temperature vs holes density for 30% of Mn diluted in ZnO containing 319 Zn atoms (small cell) and 1270 Zn atoms (big cell). We deduce that $T_C$ reaches its maximum at $n_C$=0.58, and it is higher when the cell size is bigger.

## Conclusions

The present Monte Carlo study of $Zn_{1-x}Mn_xO$ with different concentrations of Mn dopant ranging from 0.15 to 0.35 and for different values of the carrier concentrations leads to the two main results that we summarize as follows:
- The curie Temperature $T_C$ is function of many parameters as: Magnetic cation concentrations, cell size, short or long range, holes density.
- To get the FM phase with high $T_C$, the concentration of carriers should belong to the range [0.52-0.65] for the case of $Zn_{1-x}Mn_xO$.